\documentclass[]{valhalla}

\usepackage{amsmath,amsfonts,bm}

\def\eqref#1{equation~\ref{#1}}

\def\1{\bm{1}}

\DeclareMathAlphabet{\mathsfit}{\encodingdefault}{\sfdefault}{m}{sl}
\SetMathAlphabet{\mathsfit}{bold}{\encodingdefault}{\sfdefault}{bx}{n}

\usepackage{iftex}
\usepackage{url}
\usepackage{booktabs}
\usepackage{graphicx}
\usepackage{float}
\usepackage{xcolor}
\usepackage{colortbl}
\usepackage{makecell}
\usepackage{pifont}
\usepackage{wrapfig}
\usepackage{hyperref}
\usepackage{array}
\usepackage{multirow}
\usepackage{amsmath}
\usepackage{amssymb}
\usepackage{amsfonts}
\usepackage{nicefrac}
\usepackage{microtype}
\usepackage{longtable}
\usepackage{enumitem}
\usepackage{subcaption}
\usepackage{cleveref}
\usepackage[toc,page,header]{appendix}
\usepackage{minitoc}
\usepackage{tcolorbox}
\tcbuselibrary{most}

\makeatletter
\newcommand{\EpibenchAppendixHeadingSpacing}{%
    \renewcommand{\section}{\@startsection{section}{1}{\z@}{-0.8ex plus -0.2ex minus -.1ex}{0.25ex plus .08ex}{\large\sffamily\bfseries\color{valhallaprimary}\raggedright}}%
    \renewcommand{\subsection}{\@startsection{subsection}{2}{\z@}{-0.55ex plus -0.16ex minus -.08ex}{0.16ex plus .05ex}{\normalsize\sffamily\bfseries\color{valhallaprimary}\raggedright}}%
    \renewcommand{\subsubsection}{\@startsection{subsubsection}{3}{\z@}{-0.18ex plus -0.06ex minus -.04ex}{0.05ex plus .02ex}{\normalsize\sffamily\bfseries\color{valhallaprimary}\raggedright}}%
}
\newcommand{\EpibenchNeedspace}[1]{%
    \par\begingroup
    \dimen@=#1\relax
    \dimen@ii=\pagegoal
    \advance\dimen@ii by -\pagetotal
    \ifdim\dimen@>\dimen@ii
        \pagebreak
    \fi
    \endgroup
}
\makeatother

\newcounter{epialgorithm}

\definecolor{ReasoningBand}{HTML}{F1EAFE}
\definecolor{NonreasoningBand}{HTML}{F7F3EE}
\newsavebox{\exampleboxcontent}
\newsavebox{\systempromptboxcontent}
\newsavebox{\caseboxcontent}
\newlength{\exampleboxwidth}
\newlength{\systempromptboxwidth}
\newlength{\caseboxwidth}

\paperlogo{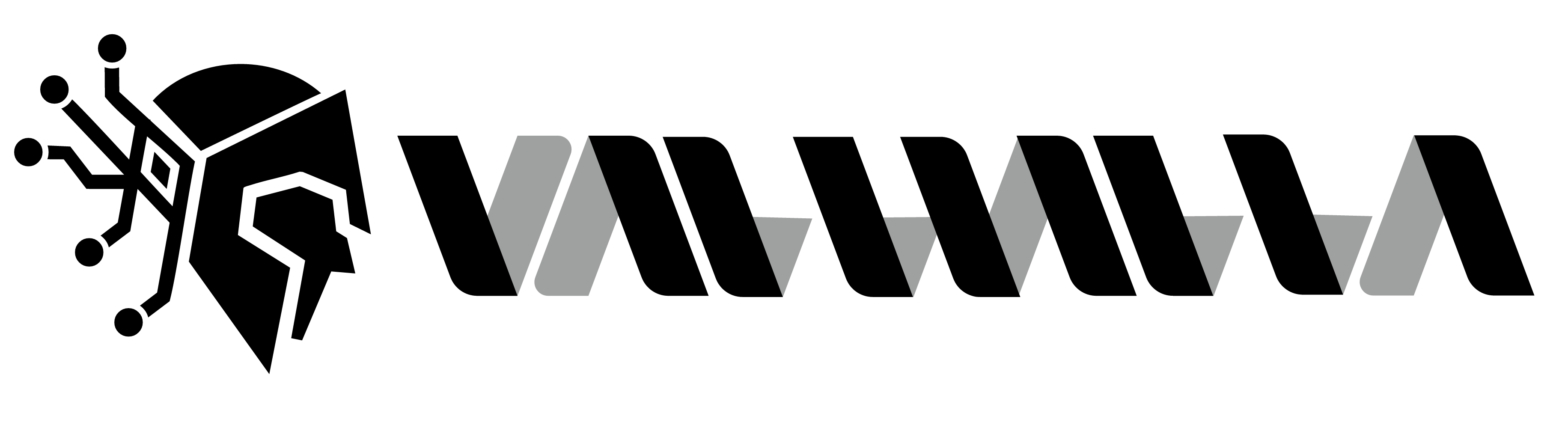}
\papertagline{[Paper tagline]}

\title{De novo design of monoclonal and bispecific antibodies with OFAntibody}
\author[*]{Valhalla Team}

\abstract{Recent advances in generative protein design have enabled de novo antibody generation with explicit target and epitope conditioning. However, most existing approaches remain formulated around a single antigen-antibody interface, whereas bispecific antibody design requires modeling multi-component complexes in which multiple target-recognition interfaces must coexist and interact within a shared antibody structure. Here we present OFAntibody, an all-atom generative framework for de novo design of monoclonal and bispecific antibodies. OFAntibody expands CDR-epitope interaction learning with large-scale distilled antigen-antibody complexes, and introduces multi-component structural supervision and arm-aware multi-hotspot routing to learn compatible multi-interface geometries and couple each antibody paratope to its designated epitope. OFAntibody supports epitope-conditioned generation across monoclonal antibodies and diverse bispecific formats, including tandem VHH, diabody and CODV. In nanobody design benchmarks, OFAntibody achieves a Top-5 enrichment rate of 41.5\%, representing a 5.39-fold improvement over RFantibody in competitive candidate ranking. In bispecific antibody design tasks, OFAntibody achieves hotspot pass rates of 94–100\% across the three evaluated tasks and achieves energy pass rates of 60\%, 13\% and 8\% for diabody, tandem VHH and CODV formats, respectively. OFAntibody further enables the same target combination to be explored across different antibody formats, while joint multi-interface generation reduces geometric incompatibilities arising from independent design and post hoc assembly. Together, these results extend de novo antibody design from binary antigen-antibody complexes to programmable multi-component complexes, providing a foundation for designing single molecules that combine recognition of distinct targets and their associated biological functions.}

\checkdata[Code]{\url{https://github.com/OTeam-AI4S/OFAntibody.git}}

\begin{document}
\maketitle

\section{Introduction}

Antibody therapeutics have become one of the most important classes of therapeutic molecules in modern biomedicine owing to their high target specificity, excellent binding affinity and engineerable molecular scaffolds~\cite{carter2018next}. Conventional monoclonal antibodies exert therapeutic functions through diverse mechanisms, including blocking ligand–receptor interactions, inducing receptor internalization and recruiting Fc-mediated immune effects. Nanobodies, derived from camelid single-domain antibodies, further expand the scope of antibody engineering owing to their small molecular size, high stability and ease of recombinant expression. Their compact architecture also enables flexible engineering strategies, including multivalent assembly~\cite{jin2023nanobodies}. However, antibody discovery and optimization remain highly dependent on large-scale experimental validation, including animal immunization~\cite{kohler1975continuous}, display-based screening~\cite{mccafferty1990phage} and affinity maturation~\cite{wang2021optimization}. The highly diverse sequence and conformational space of antibody complementarity-determining regions (CDRs) makes antigen–antibody recognition difficult to engineer computationally. Because antibody recognition depends on the complex interplay among epitope properties, interface geometry and CDR conformational states~\cite{sela2013structural}, designing antibodies with desired binding modes within the vast sequence–structure space remains a central challenge in computational antibody engineering.

Recent advances in deep learning-based protein structure prediction and generative modeling have substantially transformed protein engineering. AlphaFold3~\cite{abramson2024accurate} and other biomolecular co-folding models~\cite{chai2024chai,bytedance2025protenix,wohlwend2025boltz,evans2021protein} have significantly improved the accuracy of protein complex structure prediction. Meanwhile, advances in de novo protein design have combined generative structure modeling with structure-conditioned sequence design. Methods such as RFdiffusion~\cite{watson2023novo} generate protein backbones conditioned on structural motifs or target interfaces, while ProteinMPNN~\cite{dauparas2022robust} identifies sequences compatible with the designed structures. These developments demonstrate that generative models can navigate protein sequence–structure space under explicit design objectives,  providing a foundation for extending generative approaches to antibody systems. However, extending generative approaches to antibody systems remains challenging. Compared with general protein binder design~\cite{watson2023novo,pacesa2025one}, antibody design is more structurally constrained because the binding interface is largely confined to CDR loops presented by the immunoglobulin framework. Unlike general binder design, where the binding interface can be created more freely across a de novo generated scaffold, antibody design largely depends on redesigning CDR sequences and conformations within a constrained antibody framework. Therefore, a central challenge is to design CDRs that form precise and complementary interactions with predefined antigen epitopes.

In this context, antibody generation models have progressively advanced from local CDR optimization toward target-conditioned antibody generation. Early approaches, including DiffAb~\cite{luo2022antigen}, MEAN~\cite{kong2022conditional} and dyMEAN~\cite{kong2023end}, focused on designing CDR conformations and sequences conditioned on predefined antibody frameworks and antigen contexts, formulating antibody design as a local CDR redesign problem. RFantibody~\cite{bennett2026atomically} further advanced this paradigm by extending RFdiffusion-based generative modeling to antibody systems, enabling de novo design of antibodies targeting user-defined epitopes with experimental validation. Subsequently, IgGM~\cite{wang2025iggm} explored joint sequence–structure co-design for antibodies under predefined antigen conditions, enabling tighter integration of sequence and structural information during antibody generation. Beyond antibody-specific models, recent advances in general biomolecular binder generation have incorporated antibody design into unified generative frameworks. PPIFlow~\cite{yu2026high} introduced a flow matching framework supporting the design of protein binders, antibodies and nanobodies, whereas BoltzGen~\cite{stark2025boltzgen} and AnewOmni~\cite{kong2026programming} further advanced all-atom generative modeling across diverse molecular modalities. ODesign~\cite{zhang2025odesign} extended this paradigm toward all-to-all biomolecular interaction design involving proteins, nucleic acids and small molecules, while supporting conditional generation of antibody-related binders. Despite these advances, existing approaches remain primarily focused on single-interface binary complex generation and lack the ability to systematically design bispecific antibodies containing two coupled antigen–antibody interfaces.

Bispecific antibodies represent an increasingly important therapeutic modality by combining distinct target-recognition activities and biological functions within a single molecule~\cite{budde2022safety,steinmetz2016codv}. For example, CD20$\times$CD3 bispecific antibodies simultaneously recognize CD20-expressing target cells and recruit CD3-expressing T cells, coupling antigen recognition with immune-cell engagement~\cite{bacac2018cd20}. This functional integration, however, introduces a distinct structural design problem. Bispecific antibody design cannot be reduced to the independent design and subsequent assembly of two monospecific binding units, because the two binding interfaces are coupled through the shared antibody architecture. A successful design must position each binding arm toward its designated epitope while maintaining compatibility between the two antigen--antibody interfaces and the overall bispecific format. Domain arrangement, interdomain connectivity, linker geometry and neighboring variable domains constrain the relative positions, orientations and accessibility of the two paratopes, and may sterically restrict antigen access to either binding site~\cite{digiammarino2011ligand,jakob2013structure,steinmetz2016codv}. Consequently, two binding units that individually form favorable antigen--antibody interfaces may become geometrically incompatible when incorporated into the same bispecific molecule~\cite{lu2004effect,sato2025bispecific}. In practice, bispecific antibody development therefore still relies heavily on expert-guided format selection and iterative screening of parental binder combinations and molecular architectures, although high-throughput assembly platforms such as FORCE can accelerate this process~\cite{dengl2020format}. To complement these experimental strategies, we formulate bispecific antibody design as a format-conditioned multi-interface generation problem, in which arm-specific epitope recognition, compatibility between the two binding interfaces and the overall antibody architecture are jointly considered within a predefined bispecific template.

To address these challenges, we develop OFAntibody, a unified generative framework for monoclonal and bispecific antibody design. OFAntibody enables conditional generation of antibody backbones and epitope-specific CDR-mediated interfaces while supporting format-conditioned design of bispecific antibodies within a single framework. OFAntibody incorporates three key innovations. First, we develop an arm-aware multi-hotspot routing mechanism, which allows independent specification of hotspot sets for different targets or epitopes and explicit assignment of each hotspot set to its corresponding antibody binding arm. By encoding the structured relationship between targets, hotspots and binding arms, this mechanism enables joint control over target identities, epitope locations and binding-arm assignments, thereby reducing ambiguity during multi-interface antibody generation. Second, we construct a structural dataset of experimentally resolved ternary complexes containing three interacting molecular components, providing supervision for learning multi-interface geometric relationships. This enables the model to capture coupled spatial constraints among multiple interaction interfaces and facilitates the generation of bispecific antibody designs. Third, we expand antibody–antigen interaction data through structural distillation of antibody--antigen pairs obtained from ASD~\cite{czerwinski2026asd}, generating approximately 150,000 antibody–antigen interaction pairs. This expanded dataset substantially increases the diversity of antigen–antibody interface geometries and interaction patterns learned by the model. The resulting knowledge from abundant single-interface interactions not only improves monoclonal antibody design but also provides a foundation for the more data-scarce problem of multi-interface antibody generation.

To our knowledge, OFAntibody represents the first unified generative framework for both monoclonal and bispecific antibody design. In nanobody design benchmarks, OFAntibody achieves superior performance in competitive candidate ranking across diverse targets, with a Top-5 enrichment rate of 41.5\%, 5.39-fold higher than RFantibody. OFAntibody also improves overall candidate prioritization, achieving lower average and median rankings than RFantibody (270.3 versus 301.8 and 250 versus 300, respectively). In bispecific antibody design tasks, OFAntibody successfully satisfies predefined hotspot constraints across all evaluated cases and achieves energy pass rates of 60\%, 13\% and 8\% for diabody, tandem VHH and CODV formats, respectively. These advances establish OFAntibody as a unified generative framework that extends antibody generation beyond single-interface recognition toward format-conditioned multi-interface antibody design.
\section{Results}

\subsection{Overview of OFAntibody}

OFAntibody is a unified all-atom antibody structure design model that enables monoclonal and bispecific antibody generation conditioned on target antigen structures, epitope constraints and bispecific antibody formats (Fig. 1a). The model takes target antigen structures together with antibody design constraints and conditioning information as input, and generates antibody–antigen complex conformations satisfying the specified structural requirements. For monoclonal antibody design, OFAntibody generates CDR conformations on predefined VHH or ScFv frameworks. For bispecific antibody design, OFAntibody generates designs under predefined bispecific antibody formats, including tandem VHH, diabody and CODV, by jointly modeling CDR conformations from different binding arms and the conformations of connecting regions.

The generative framework of OFAntibody incorporates three design principles: bispecific antibody format conditioning, arm-aware multi-hotspot routing and training with multi-source antibody structural data (Fig. 1). The format-conditioning mechanism enables the model to adapt to diverse bispecific antibody formats and structural organizations. The arm-aware multi-hotspot routing mechanism explicitly encodes the correspondence between target epitopes, hotspot regions and antibody binding arms, enabling structured conditioning of multiple epitope constraints (Fig. 1c). The training dataset comprises experimentally resolved antibody–antigen complexes, ternary complex structures and distilled antibody–antigen interaction data (Fig. 1b,d,e), covering both single-interface antibody recognition patterns and the spatial organization of multiple interacting interfaces.

Based on this framework, OFAntibody supports diverse antibody modalities, including nanobodies and paired ScFvs, as well as multiple bispecific antibody formats, including tandem VHHs, diabodies and CODV. The model generates antibody structures conditioned on target epitopes, antibody structural constraints and binding-arm assignments, enabling controllable design across different antibody systems. We subsequently evaluate OFAntibody in three design settings: nanobody design, multi-format bispecific antibody design and joint multi-interface antibody design.

\begin{figure}[htbp]
    \centering
    \includegraphics[
        width=\linewidth,
        page=2
    ]{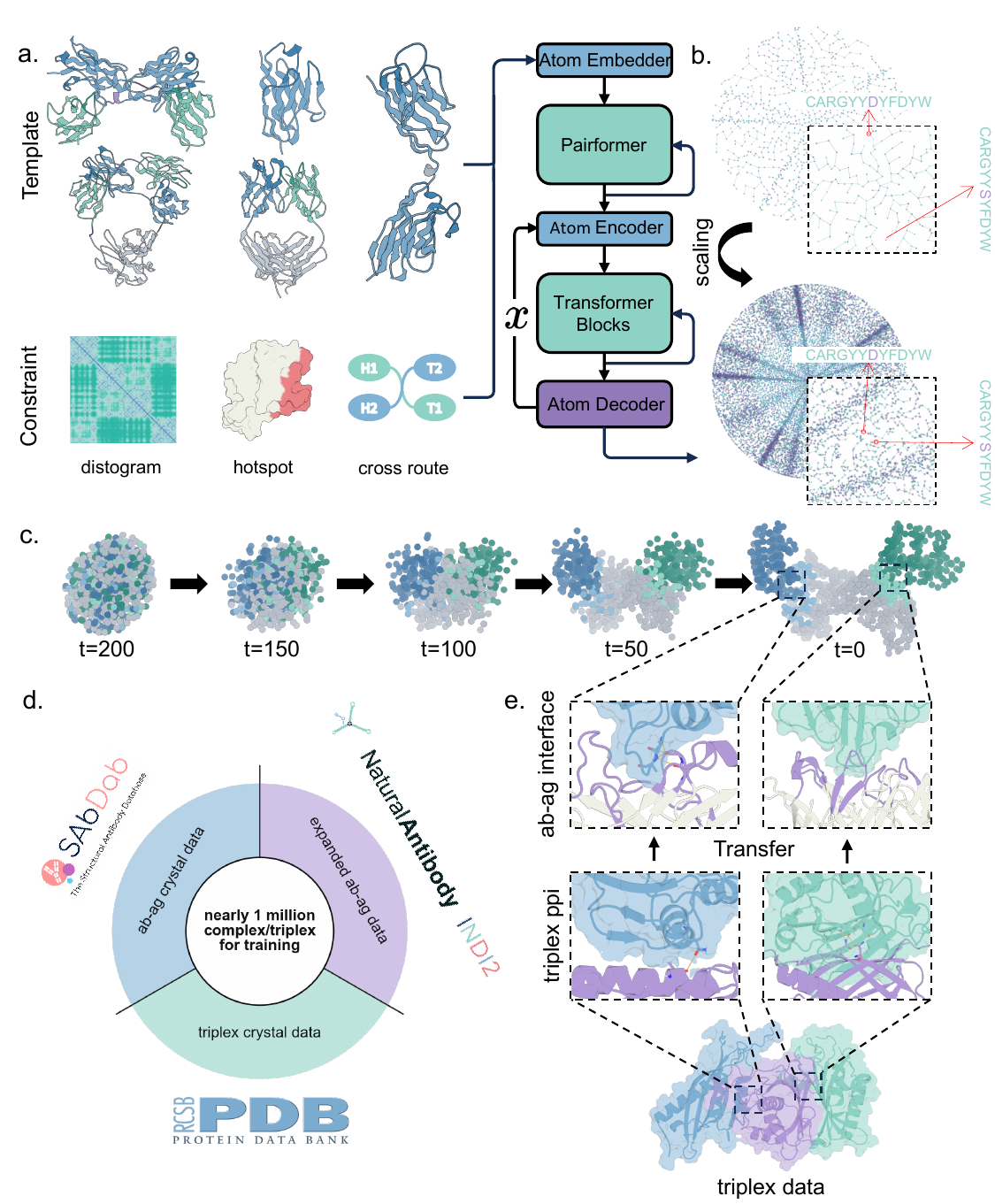}
    \caption{Overview of the OFAntibody framework for monoclonal and bispecific antibody design. a, Overview of the OFAntibody workflow and conditioning strategies. OFAntibody generates monoclonal and bispecific antibody structures under format, structural and epitope constraints. b, Expansion of the antibody–antigen interaction space through distillation data, increasing the diversity of learned interface geometries and interaction patterns. c, Arm-aware multi-hotspot routing mechanism for bispecific antibody design. Target epitopes and corresponding binding arms are independently specified, and routing constraints guide each binding arm toward its assigned epitope to enable predefined binding-arm–target mapping. d, Composition of the OFAntibody training dataset, including antibody–antigen complexes, ternary complex structures and distilled antibody–antigen interaction data. e, Integration of antibody–antigen interaction knowledge for ternary complex modeling. Monoclonal antibody data provides interface interactions and CDR geometries, while ternary complex data provides additional spatial constraints for multi-interface modeling.}
    \label{fig:1}
\end{figure}

\subsection{OFAntibody achieves state-of-the-art in silico performance in nanobody design}

To systematically evaluate the nanobody design performance of OFAntibody, we compared it with representative antibody and protein binder design approaches, including IgGM~\cite{wang2025iggm}, BoltzGen~\cite{stark2025boltzgen}, PPIFlow~\cite{yu2026high}, RFantibody~\cite{bennett2026atomically} and AnewOmni~\cite{kong2026programming}. Unlike evaluations that assess candidates generated by each method independently, we adopted a competitive ranking strategy in which candidates from all methods were evaluated jointly under a unified framework. This setting better reflects practical experimental screening scenarios, where only a limited number of candidates can be experimentally characterized and designs generated by different approaches compete for limited experimental validation opportunities.

We constructed a nanobody design benchmark comprising 13 design tasks across 12 targets, covering diverse antigen classes, epitope constraints and VHH structural frameworks. The benchmark includes viral antigens (SARS-CoV-2 RBD, SARS-CoV-2 Core and influenza HA), soluble inflammatory factors (TNF$\alpha$, TSLP and IL-33), therapeutic targets associated with cancer immunotherapy (HER2, EGFR and PD-L1), a coagulation-related target (VWF-A1), as well as structurally diverse membrane proteins and immune receptors (CCR8, NKp30 and LGR4). For each design task, predefined hotspot constraints and four VHH structural frameworks were evaluated to assess model performance across diverse target environments, epitope conditions and antibody scaffold geometries.

Before candidate ranking, we first performed Geometry Check to remove generated structures containing chain breaks or severe steric clashes. Candidates passing geometric filtering were further evaluated using structural confidence metrics and interface physical properties. Following the worst-rank aggregation strategy introduced by BoltzGen~\cite{stark2025boltzgen}, each metric was converted into a relative ranking, and the worst metric rank among all evaluated criteria was used as the final candidate score. We then quantified the enrichment of each method among the Top-5, Top-10 and Top-20 candidates, together with the average and median candidate rankings. Detailed sampling procedures, geometric filtering criteria and ranking metrics are described in the Methods.

OFAntibody achieved superior performance in competitive candidate ranking across the benchmark. Across all 13 design tasks, OFAntibody-generated candidates achieved enrichment rates of 41.5\%, 39.2\% and 36.5\% among the Top-5, Top-10 and Top-20 candidates, respectively, representing the highest enrichment across all evaluated methods (Fig. 2a). In the Top-10 evaluation, OFAntibody achieved the highest enrichment ratio in 11 of the 13 tasks, including LGR4, a target with distinct structural and epitope characteristics. These results suggest that OFAntibody maintains strong design performance across both well-characterized antibody targets and targets with diverse structural and epitope properties. Consistent with the Top-K enrichment analysis, OFAntibody also achieved improved overall candidate prioritization, with a median ranking of 250.0, 28.5 lower than the second-best BoltzGen, and an average ranking of 270.3, 18.0 lower than the second-best IgGM (Fig. 2c).

To understand the factors underlying OFAntibody performance, we analyzed the distribution of individual ranking metrics across different methods (Fig. 2b). For structural confidence metrics, OFAntibody-generated complexes exhibited higher iPTM scores and lower minimum interface PAE values, indicating higher confidence in the predicted antigen–antibody interfaces and relative inter-chain arrangements. In addition, higher binder pTM scores indicated improved structural confidence of the generated VHH molecules. For interface physical properties, OFAntibody designs generally formed more hydrogen bonds and salt bridges and exhibited larger $\Delta$SASA values, suggesting more extensive polar interactions and interface burial. Notably, most candidates generated by AnewOmni failed Geometry Check due to chain discontinuities or severe steric clashes and were therefore excluded from enrichment analysis. Although a subset of AnewOmni candidates exhibited favorable local interface metrics, including hydrogen bond numbers, salt bridges and $\Delta$SASA, these structures also frequently displayed excessive penetration into target surfaces or atomic overlaps, together with lower AF3 prediction confidence. These observations indicate that local interface contact metrics alone are insufficient to evaluate antibody design quality and should be interpreted together with global structural validity and interface confidence.

Finally, we investigated the contribution of distilled antibody–antigen interaction data to OFAntibody performance (Fig. 2d). Compared with OFAntibody trained without distillation data, the complete model improved Top-5, Top-10 and Top-20 enrichment rates by 9.2, 6.9 and 7.3 percentage points, respectively, with the largest improvement observed for Top-5 enrichment. These results suggest that large-scale distilled interaction data improves the model’s ability to learn diverse antigen–antibody interface geometries and interaction patterns, thereby enhancing nanobody design performance across diverse targets and epitope conditions.

\begin{figure}[htpb]
    \centering
    \includegraphics[
        width=\linewidth,
        page=3
    ]{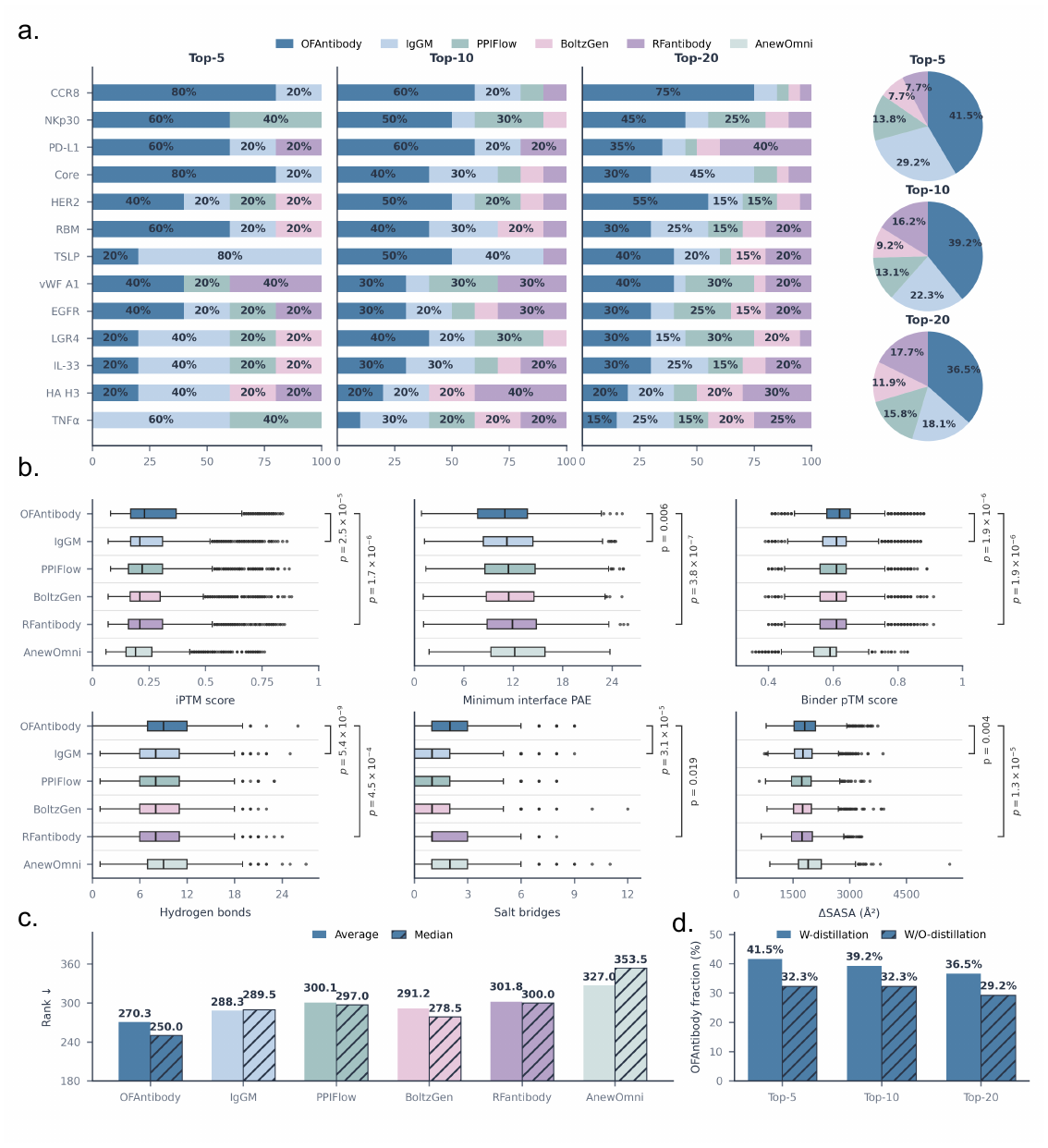}
    \caption{OFAntibody enables competitive candidate ranking for nanobody design. a, Competitive enrichment of OFAntibody and baseline methods across 13 nanobody design tasks. Stacked bar plots show the contribution of each method among the Top-5, Top-10 and Top-20 ranked candidates for each target. Pie charts summarize the overall enrichment across all design tasks. b, Distribution of ranking metrics used for candidate prioritization, including iPTM, minimum interface PAE, binder pTM, hydrogen bonds, salt bridges and $\Delta$SASA. Statistical significance is indicated as shown. c, Average and median candidate rankings generated by different methods. Lower values indicate better overall candidate prioritization. d, Ablation of antibody–antigen interaction distillation by comparing OFAntibody models trained with and without distilled interaction data based on Top-5, Top-10 and Top-20 enrichment rates.}
    \label{fig:2}
\end{figure}

\subsection{OFAntibody enables multi-format bispecific antibody design under predefined epitope constraints}

Building on its capability in monoclonal antibody design, OFAntibody extends antibody generation toward multi-interface bispecific antibody design. Unlike monoclonal antibody design, where a single antigen–antibody interface is optimized, bispecific antibody design requires joint optimization of two coupled binding interfaces and their compatibility with a predefined antibody format. To evaluate the multi-interface design capability of OFAntibody, we selected representative bispecific antibody design tasks and assessed generated candidates from four aspects: epitope targeting, CDR-mediated interactions, structural compatibility and interface energetics. Specifically, we defined four evaluation metrics (detailed thresholds and calculation procedures are provided in Methods): (1) Hotspot Pass Rate, defined as the proportion of designs in which both binding interfaces satisfy predefined hotspot constraints; (2) CDR Interaction Pass Rate, defined as the proportion of designs in which both interfaces satisfy criteria describing canonical CDR-mediated interactions; (3) Geometry Pass Rate, defined as the proportion of designs that simultaneously satisfy hotspot and CDR interaction criteria without severe atomic clashes; and (4) Energy Pass Rate, defined as the proportion of designs that exhibit more favorable computed interface energies than the corresponding reference binders at both interfaces.

To evaluate OFAntibody under diverse bispecific antibody format constraints, we selected three representative formats: tandem VHH (VHH–linker–VHH), diabody and CODV--like~\cite{steinmetz2016codv} formats (Fig. 3a). These formats impose distinct geometric constraints through differences in domain organization, chain connectivity and relative binding-arm orientation. Tandem VHH consists of two independently folded VHH domains connected by a flexible peptide linker, where the relative orientation of the two binding arms is primarily determined by linker length and flexibility (Fig. 3a, middle). The single-chain diabody format used in this study contains four VH/VL domains arranged within one polypeptide chain, where non-adjacent VH and VL domains pair to form two Fv binding sites, resulting in a compact bispecific format (Fig. 3a, left). CODV-like molecules consist of two multi-domain chains, in which two VH/VL pairs form crossed binding sites. The C-terminal CH1 and CL constant domains further stabilize assembly of the two chains, resulting in a compact triangular molecular organization (Fig. 3a, right). Together, these three formats provide a stringent evaluation of OFAntibody’s ability to jointly model binding interfaces under distinct bispecific antibody format constraints.

Under this multi-format evaluation framework, we selected three representative design tasks: EGFR×CD3, CD28×CD3 and EGFR 7D12×EgA1. EGFR×CD3 was designed in a diabody format using the recently resolved bispecific antibody–target complex structure PDB 9IP8~\cite{sato2025bispecific} as a reference. CD28×CD3 was designed using a CODV-like format based on PDB 6O8D~\cite{wu2020trispecific}. EGFR 7D12×EgA1~\cite{schmitz2013structural} was designed in a tandem VHH format because the spatial distance and relative orientation of the two EGFR epitopes are compatible with tandem VHH topology, and experimentally characterized VHH binders 7D12 and EgA1 provided reference binding modes for evaluation. As shown in Table~1, OFAntibody-generated designs achieved Hotspot Pass Rates of 94\% for the EGFR$\times$CD3 task and 100\% for the other two tasks. Across all three tasks, the designs covered an average of 96.79\% of predefined hotspot residues. The CDR Interaction Pass Rates reached 69\%, 10\% and 68\% for EGFR×CD3, CD28×CD3 and EGFR 7D12×EgA1, respectively. At the interface energy level, OFAntibody achieved Energy Pass Rates of 60\%, 8\% and 13\% across the three tasks, respectively, and all generated designs exhibited negative Rosetta interface energies (Fig. 3b). These results indicate that OFAntibody can generate bispecific antibody designs under predefined epitope–binding-arm assignments while maintaining CDR-mediated interface interactions and favorable computed interface energetics.

\begin{table}[htbp]
\centering
\caption{Bispecific antibody design benchmark.}
\label{tab:bispecific_benchmark}

\begin{tabular}{llccccc}
\toprule
Task & Head & N & Hotspot & CDR Int. & Geometry & Energy \\
\midrule

\multirow{3}{*}{EGFR$\times$CD3}
& Overall &100&0.94&0.69&0.65&0.60\\
& EGFR&100&0.94&0.70&0.66&0.67\\
& CD3&100&1.00&0.89&0.89&0.87\\

\midrule

\multirow{3}{*}{CD28$\times$CD3}
& Overall&100&1.00&0.10&0.10&0.08\\
& CD28&100&1.00&1.00&1.00&1.00\\
& CD3&100&1.00&0.10&0.10&0.08\\

\midrule

\multirow{3}{*}{EGFR 7D12$\times$EgA1}
& Overall&100&1.00&0.68&0.68&0.13\\
& 7D12&100&1.00&0.68&0.68&0.15\\
& EgA1&100&1.00&1.00&1.00&0.88\\

\bottomrule
\end{tabular}

\end{table}

We next investigated the lower CDR Interaction Pass Rate observed in the CD28×CD3 CODV-like task. The CDR Interaction Pass Rate for this task was only 10\%, substantially lower than the other two tasks, with the major limitation originating from the CD3 binding arm (Table 1). To determine whether this reflected insufficient CDR-mediated interactions, we analyzed polar interactions formed at the designed CD3 interface. Among all hydrogen bonds and salt bridges, 98.9\% and 91.5\%, respectively, were mediated by CDR residues (Fig. 3c), indicating that CDR residues still contributed the majority of key interface interactions. The discrepancy between the high contribution of CDR residues and the low CDR Interaction Pass Rate suggests that criteria derived from canonical antibody interaction patterns may not fully capture interface geometries in CODV-like formats. The unique triangular domain organization of CODV may redistribute interface contacts among different CDR regions, allowing some designs to form CDR-dominated and structurally plausible interactions while failing to satisfy predefined residue-number and distance thresholds. Therefore, this result primarily reflects limitations of current CDR interaction metrics in evaluating diverse bispecific antibody geometries and highlights the need for format-specific interface evaluation criteria.

Beyond CDR interaction criteria, we further analyzed the relatively lower Energy Pass Rate observed in the CD28×CD3 CODV-like task. By separately evaluating the two binding interfaces, we found that the major limitation originated from the CD3 interface, where only 8\% of designs achieved interface energies more favorable than the reference binder (Table 1). Compared with the OKT3–CD3 reference interface, OFAntibody designs reproduced 46.4\% of hydrogen bonds and 87.2\% of salt bridges observed in the reference structure (Fig. 3d), indicating that the model preserved a substantial fraction of reference polar interactions under the geometric constraints imposed by the CODV format. Meanwhile, newly generated hydrogen bonds and salt bridges were also observed, suggesting that OFAntibody can explore alternative local interaction patterns while satisfying constrained bispecific antibody geometries.

\begin{figure}[htbp]
    \centering
    \includegraphics[
        width=\linewidth,
        page=4
    ]{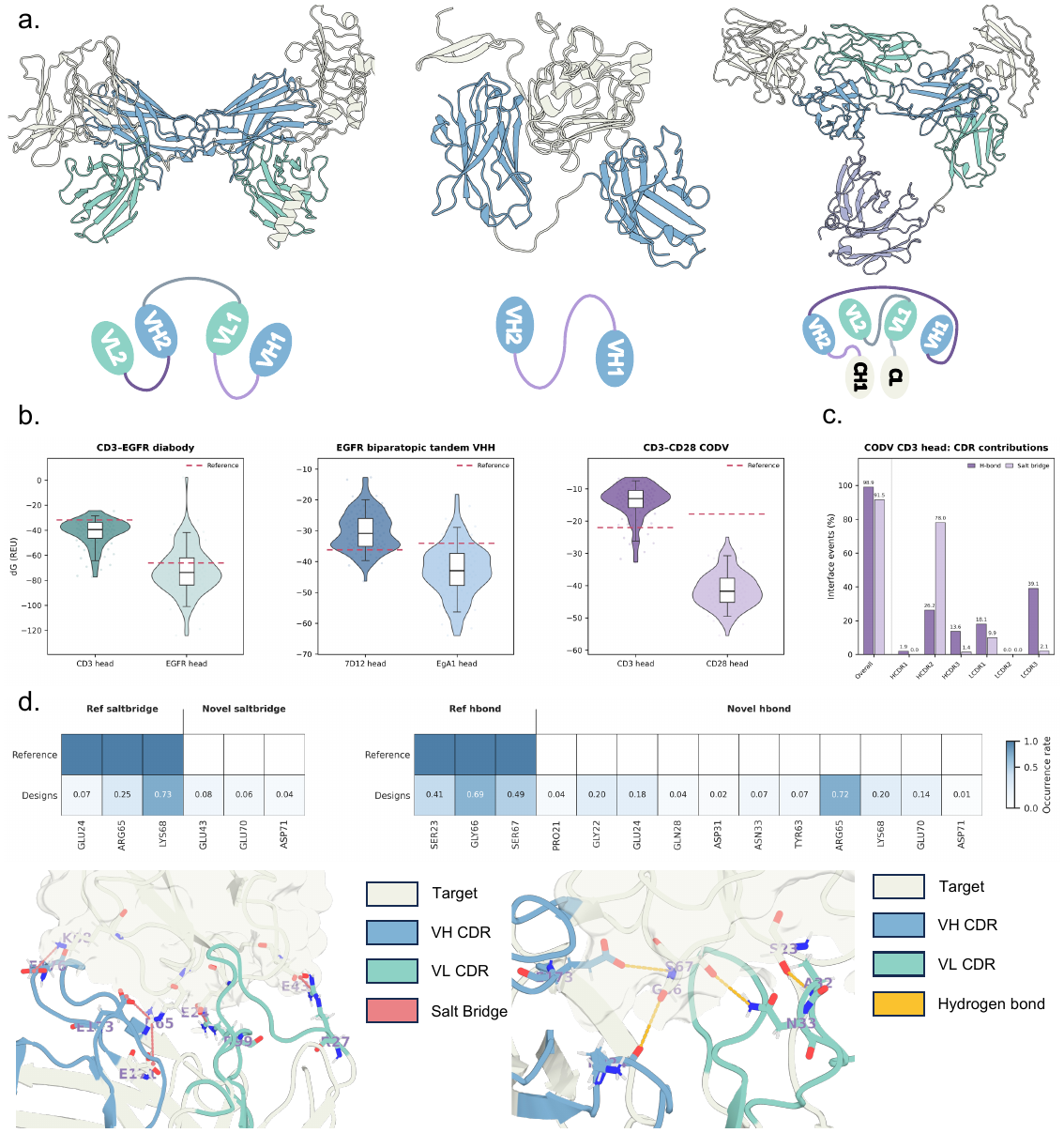}
    \caption{OFAntibody enables multi-format bispecific antibody design. a, Three bispecific antibody formats evaluated in this study: diabody, tandem VHH and CODV. The upper panels show the three-dimensional structures of the corresponding formats, and the lower panels show their two-dimensional schematic representations. b, Interface energy distributions of OFAntibody-generated designs across three bispecific antibody design tasks: EGFR×CD3 diabody, EGFR 7D12×EgA1 tandem VHH and CD28×CD3 CODV. Red dashed lines indicate the interface energies of the corresponding reference complexes. c, Contribution of different CDR regions to polar interactions at the CD3-binding arm in the CD28×CD3 CODV design task. d, Interaction analysis of the CD3-binding interface in CD28×CD3 CODV designs. Heatmaps show the occurrence frequency of reference and newly formed salt bridges and hydrogen bonds across generated designs. Representative CDR-mediated salt bridges and hydrogen bonds are shown below.}
    \label{fig:3}
\end{figure}

\subsection{Joint design of bispecific antibodies outperforms independent design and assembly of monoclonal binders}

Bispecific antibody design cannot be reduced to the simple combination of two independently optimized monospecific binding modules. Even when individual binding interfaces exhibit high structural confidence and favorable computed energetics, their combination may fail to form geometrically compatible bispecific assemblies due to constraints imposed by relative domain orientation, spatial accessibility and linker connectivity. To investigate whether joint multi-interface design improves overall bispecific compatibility, we compared two design strategies under identical epitope constraints: joint design and independent binder design followed by assembly. In the joint design strategy, two binding arms and their connecting topology are optimized simultaneously, whereas in the independent assembly strategy, two binding arms are first designed separately against their respective epitopes and subsequently combined based on their individual binding conformations.

We used a tandem VHH targeting the NM02 and 2RS15d HER2 epitopes~\cite{sawmynaden2023co,d2017131i} as a representative case to compare these two strategies. In the joint design setting, hotspot constraints corresponding to the two HER2 epitopes were simultaneously specified, and OFAntibody jointly generated the CDR conformations of both VHH domains together with the linker conformation under a predefined tandem VHH format. In the independent assembly strategy, two VHH binders targeting the two HER2 epitopes were independently designed under the same epitope constraints. High-confidence monospecific binders were selected based on individual binding predictions, followed by VHH combination, linker construction and interface geometry analysis. Detailed sampling procedures and evaluation criteria for both strategies are described in Methods.

The two strategies exhibited substantial differences in geometric compatibility. Among the 2,130 combinations generated by independent binder assembly, 1,708 combinations (80.2\%) exhibited severe VHH–VHH steric clashes (Fig. 4b). After removing these incompatible combinations, an additional 246 combinations were excluded because their relative orientations or inter-domain distances prevented effective linker connection. Ultimately, only 176 combinations proceeded to linker construction and optimization. In contrast, candidates generated through joint multi-interface design showed no severe spatial conflicts and maintained compatible overall conformations. These results indicate that high-quality single-interface designs do not necessarily translate into globally compatible bispecific geometries. Instead, joint generation enables simultaneous coordination of binding interfaces, binding-arm orientations and linker connectivity during the design process.

For candidates passing geometric filtering, we further compared the interface properties generated by the two strategies. Contacting residues were defined as residues containing heavy atoms within 5 Å of the target antigen. Under this criterion, CDR residues in jointly designed bispecific antibodies accounted for 77.7\% of contacting residues, representing an 11 percentage-point increase compared with independent binder assembly. Furthermore, joint design achieved 98.6\% hotspot residue coverage, 39.6 percentage points higher than independent assembly (Fig. 4c). Decomposition of the two binding interfaces revealed that independently assembled designs showed reduced hotspot geometry compatibility at the NM02 head and reduced CDR geometry compatibility at the 2RS15d head (Fig. 4c). These results suggest that independent binder assembly may compromise paratope–epitope compatibility when adapting individually optimized binding modules to a predefined tandem VHH format. In contrast, joint multi-interface generation more effectively coordinates the geometric constraints between individual binding interfaces and the overall bispecific antibody format, enabling improved preservation of interface geometry and epitope targeting.

\subsection{Exploring bispecific antibody formats as design variables using OFAntibody}

Bispecific antibody formats define molecular valency, domain organization and conformational constraints between binding arms, thereby shaping the accessible structural space for multi-interface antibody design. Consequently, even under identical target combinations and epitope constraints, different bispecific antibody formats may impose distinct geometric constraints on the formation of compatible binding interfaces~\cite{mullin2024applications,brinkmann2017making,dengl2020format}. Conventional bispecific antibody development typically identifies monospecific parental binders first and subsequently adapts them to selected formats. However, because different formats rely on distinct domain organizations and spatial arrangements, binders optimized for one format may not be directly transferable to another. This modular strategy limits systematic exploration of format-dependent structural design spaces under fixed target and epitope conditions. In contrast, OFAntibody incorporates bispecific antibody formats as generation conditions, allowing binding-arm and connecting-region conformations to be generated under different formats while maintaining the same target and epitope constraints. This enables antibody format itself to become an explicitly explored variable within the bispecific antibody design space.

To evaluate whether OFAntibody can generate compatible bispecific antibody designs under different formats while keeping target and epitope constraints fixed, we selected CD20×CD3 as a representative target combination and generated designs using three bispecific antibody formats: tandem VHH, diabody and CODV. CD20×CD3 represents a canonical T cell engager system~\cite{bacac2018cd20}, in which simultaneous recognition of tumor cells and T cells enables T cell-dependent cytotoxicity. For each format, we sampled 20,000 candidate structures and generated four candidate sequences for each structure using ProteinMPNN. AlphaFold 3 was subsequently used to predict the two individual antigen–binding-arm subcomplexes for each designed sequence, and the average iPTM score across the two interfaces was used as an overall interface-level confidence metric. All three formats generated candidates with high predicted interface confidence (mean iPTM > 0.65), with the highest mean iPTM reaching 0.72 for the diabody format. Despite the distinct binding-arm organizations and spatial constraints imposed by different formats, the top-ranked candidates from each format maintained an average hotspot coverage of 91.7\% (Fig. 4a). These results indicate that OFAntibody can preserve predefined epitope targeting while adapting antibody generation to different bispecific formats, enabling systematic exploration of format-dependent structural design spaces.

\begin{figure}[htbp]
    \centering
    \includegraphics[
        width=\linewidth,
        page=5
    ]{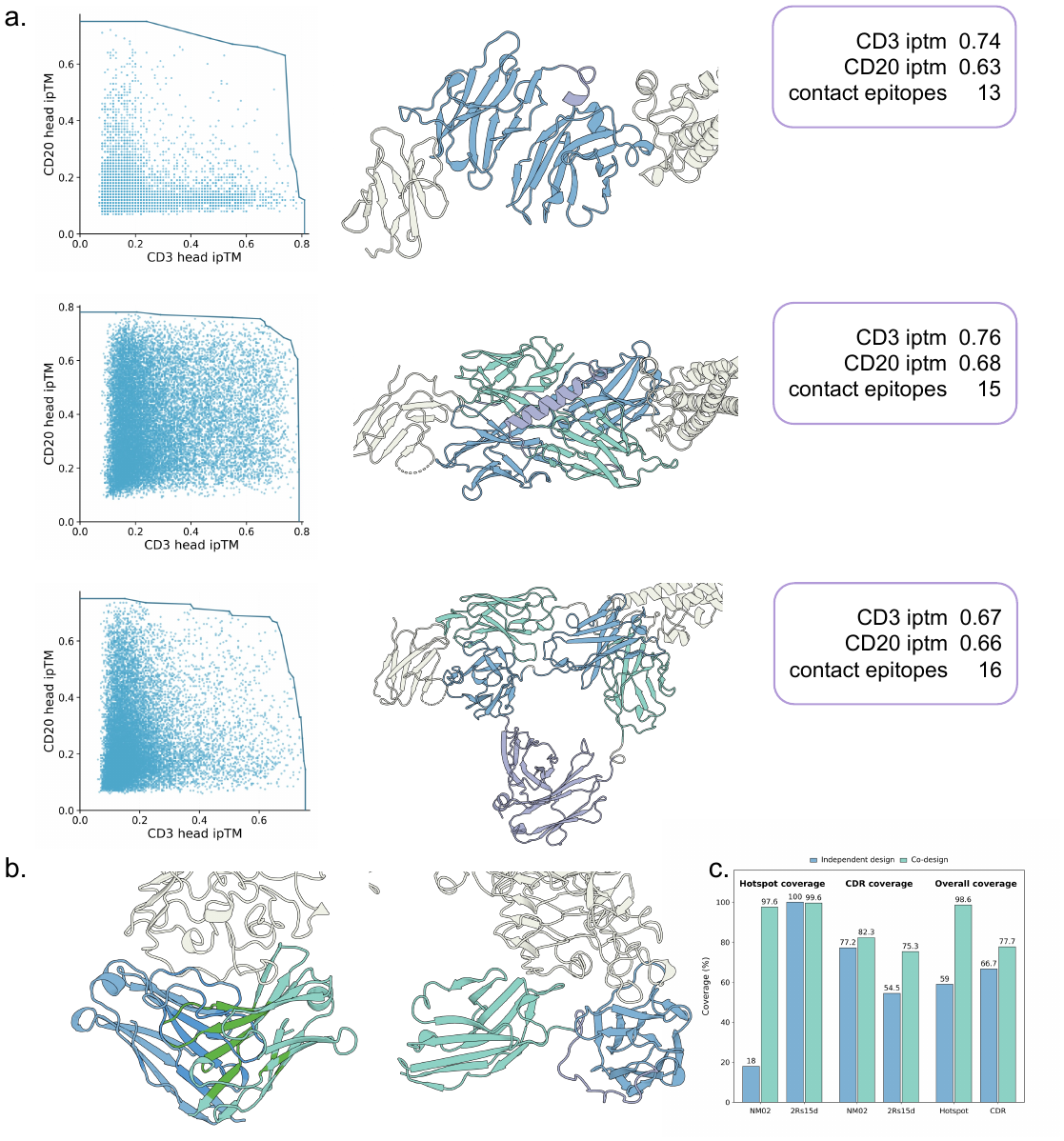}
    \caption{Multi-format and joint multi-interface design of bispecific antibodies with OFAntibody. a, OFAntibody-generated designs in three bispecific antibody formats: tandem VHH, diabody and CODV, arranged from top to bottom. For each format, the CD20 and CD3 binding interfaces are shown from left to right, together with interface iPTM scores, representative candidate structures, and corresponding interface iPTM values and hotspot coverage. b, Representative independently assembled VHHs exhibiting steric clashes (left) and a jointly designed tandem VHH complex (right).. c, Comparison of CDR coverage and hotspot residue coverage between independent binder assembly and joint multi-interface design.}
    \label{fig:4}
\end{figure}


\section{discussion}

In this work, we establish OFAntibody as a unified generative framework that extends antibody design beyond single-interface optimization toward controllable multi-interface antibody generation. By formulating antibody design as a conditional generation problem governed by structural constraints, epitope constraints and binding relationships, OFAntibody enables antibody structural constraints, target epitopes and bispecific antibody formats to be incorporated as controllable design variables within a unified framework. This formulation allows generative models to move beyond optimizing individual antigen–antibody interfaces and toward designing antibody systems involving multiple binding arms and complex engineering constraints. More broadly, our results demonstrate the potential of structure-based generative models to provide a unified computational framework for exploring diverse antibody designs that are difficult to access through conventional experimental engineering workflows.

For monoclonal antibody design, OFAntibody demonstrates that structural supervision and large-scale antibody–antigen interaction data can guide the exploration of antibody sequence–structure space under predefined epitope constraints. The distilled antibody–antigen interaction data expands the diversity of learned interface geometries and interaction patterns, providing a foundation for designing antibodies across diverse targets and binding modes. These results suggest that generative models can complement traditional antibody discovery pipelines by enabling structure-guided exploration of candidate molecules. Beyond monoclonal antibody design, OFAntibody extends the same generative framework to bispecific antibody systems, where multiple binding interfaces and molecular formats must be jointly considered. Rather than treating bispecific antibodies as assemblies of independently optimized binding modules, OFAntibody enables the joint exploration of binding interfaces, bispecific antibody formats and global molecular geometry under unified structural constraints. This capability demonstrates that molecular formats and spatial organization can be incorporated as explicit design variables, providing a scalable strategy for designing complex antibody molecules.

Despite these advances, several limitations remain. First, current evaluations rely primarily on in silico structural confidence metrics and computed interface properties, which cannot fully replace experimental validation of binding activity, specificity and developability. Future integration of experimental screening and feedback-driven optimization will be important for further improving design reliability. Second, the current framework mainly focuses on compact antibody formats, such as nanobodies and antibody fragments, whereas more complex full-length bispecific antibody systems and broader multi-interface conformational landscapes remain to be explored. Finally, the limited availability of experimentally resolved multi-interface antibody structures may constrain the model’s ability to learn complex antibody spatial organizations. Future efforts combining larger-scale structural datasets, experimental feedback loops and more comprehensive antibody engineering representations may further advance AI-driven antibody design.
\section{Methods}

\subsection{Training datasets and preprocessing}

\subsubsection{Datasets}

OFAntibody was trained on three types of datasets: antibody--antigen complex structures from SAbDab, ternary complex structures from PDB triplex, and antibody--antigen complex distillation data. Antibody--antigen crystal structures were obtained from SAbDab~\cite{dunbar2014sabdab}. A temporal split was adopted, where the training set consisted of structures released before August 16, 2024, and the test set consisted of structures released between August 16, 2024 and October 15, 2025. The PDB triplex dataset consists of preprocessed PDB ternary complex crystal structures. The training set was collected up to September 30, 2021, while the test set consisted of PDB crystal structures released between May 1, 2022 and January 12, 2023. Structures appearing in SAbDab were removed from both the training and test sets. The antibody--antigen complex distillation dataset was exclusively used as training data. All test sets were filtered to remove structures with sequence similarity greater than or equal to 40\% to any structure in the training set. For the monoclonal antibody version of OFAntibody, the PDB triplex dataset was not used, and the training dataset ratio between SAbDab and distillation data was set to SAbDab:distillation = 1:1. For the bispecific antibody version, the PDB triplex dataset was incorporated, with the training dataset ratio set to PDB triplex:SAbDab:distillation = 2:1:1. To prevent biased learning caused by imbalanced data distributions, the training datasets were clustered, and sampling was performed according to cluster categories.

\subsubsection{Distillation of antibody and nanobody complex data}

The antibody--antigen complex distillation dataset was constructed from the Antigen Specific Antibody Database (ASD DB). Antibody--antigen complexes overlapping with SAbDab were removed before further processing. Multiple sequence alignments (MSAs) were searched for all antigen--antibody pairs using ColabFold~\cite{mirdita2022colabfold}. AlphaFold2-Multimer~\cite{evans2021protein} was subsequently used to predict five structures for each antigen--antibody pair, and structures with ipTM $\geq 0.7$ were retained to construct the antibody--antigen complex distillation dataset. During dataset construction, some antigen targets were found to contain multiple domains and long disordered regions. For these targets, AFragmenter~\cite{verwimp2025afragmenter} was used to split antigen sequences based on the predicted aligned error (PAE) metric, generating domain-level antigen--antibody pairs. Each domain--antibody pair was evaluated, and the pair with the highest ipTM score was selected. Only domain--antibody pairs with ipTM $\geq 0.7$ were retained.

\subsubsection{Acquisition of the triplex dataset}

The triplex dataset was constructed by identifying ternary complexes containing one binder chain and two target chains from PDB structures. Let $\mathrm{PDBID}(c)$ denote the PDB entry corresponding to chain $c$, $\mathrm{Entity}(c)$ denote the corresponding entity identifier, $\mathcal{P}$ denote the protein set, and $D_{c_a,c_b}$ denote the distance between chains $c_a$ and $c_b$. Each candidate triplex complex was represented as:

\begin{equation}
\tau=(c_i,\{c_j,c_k\}),
\end{equation}

where $c_i$ denotes the binder chain and $\{c_j,c_k\}$ denotes an unordered pair of target chains.

A candidate triplex complex was included when the three chains originated from the same PDB entry, the binder chain corresponded to a protein entity different from both target chains, and both binder--target chain distances satisfied $D_{c_i,c_j}\leq5$ Å and $D_{c_i,c_k}\leq5$ Å. All candidates satisfying these criteria were collected as the initial triplex dataset.

To remove redundant structures originating from the same PDB entry, candidate triplex complexes were compared pairwise. Two triplex complexes were considered to have identical entity compositions when their binder chains corresponded to the same entity and their two target chains corresponded to the same set of entities regardless of target-chain ordering. For complexes with identical entity compositions, structural alignments were performed under both possible target-chain correspondence mappings, and the resulting complex RMSDs were calculated after alignment. Two triplex complexes were considered redundant when at least one correspondence mapping resulted in a complex RMSD below 2 Å.

The redundancy criteria were applied to all candidate triplex complexes, and the final triplex dataset was obtained after removing redundant samples.

\subsection{Conditioning strategies for antibody and bispecific antibody design}

\subsubsection{Sampling of multi-route hotspot residues}

During training, binder and candidate target chains are first identified at the chain level, followed by independent hotspot sampling from the binding interface of each target chain. For a complex containing $N$ residue-level tokens, each token is associated with a coordinate, chain identifier, and condition indicator. The design indicator is defined as the binary inverse of the condition indicator, where tokens with design indicators are considered designable regions.

Binder chains are determined through chain-level aggregation. Specifically, any chain containing at least one design token is considered a binder chain. For triplex structures, binder chains explicitly annotated in the dataset are directly used. All tokens belonging to selected binder chains are subsequently treated as binder tokens.

Hotspot conditioning is randomly activated during training with probability $\rho = 1/3$. When hotspot conditioning is disabled, all hotspot and binder-arm class labels are set to zero. When activated, non-binder chains within a predefined distance threshold $R = 8\,\text{\AA}$ from binder tokens are selected as candidate target chains. For each candidate target chain, hotspot residues are independently sampled from its interface region.

The number of sampled hotspot residues is independently determined for each target chain. The sampling process avoids mixing hotspot constraints from different target interfaces and ensures that each target chain receives an independent hotspot class, enabling subsequent assignment between hotspot regions and corresponding binder arms.

\subsubsection{Hotspot class and binder-arm class assignment}

To establish the correspondence between target epitopes and antibody binding arms, hotspot residues are assigned with class labels according to their target chains, and these labels are subsequently propagated to binder tokens based on spatial proximity.

Let $\mathcal{H}_c$ denote the sampled hotspot set on target chain $c$. The set of target chains containing sampled hotspots is defined as:

\begin{equation}
\mathcal{C}_{H}=\{c|\mathcal{H}_c\neq\emptyset\}.
\end{equation}

Continuous class IDs are assigned to target chains according to their ordering:

\begin{equation}
g:\mathcal{C}_{H}\rightarrow \{1,\dots,|\mathcal{C}_{H}|\}.
\end{equation}

Each hotspot token inherits the class label of its corresponding target chain:

\begin{equation}
l_j^{H}=
\begin{cases}
g(c_j), & j\in\mathcal{H},\\
0, & j\notin\mathcal{H},
\end{cases}
\end{equation}

where $\mathcal{H}=\bigcup_{c\in\mathcal{C}_{H}}\mathcal{H}_c$.

For each binder token $i\in\mathcal{B}$, the nearest sampled hotspot is identified:

\begin{equation}
j^*(i)=\arg\min_{j\in\mathcal{H}}
\|\mathbf{x}_i-\mathbf{x}_j\|_2 .
\end{equation}

If the nearest hotspot is within distance threshold $R$, the binder token inherits the corresponding hotspot class label:

\begin{equation}
l_i^{B}=
\begin{cases}
l_{j^*(i)}^{H},
& i\in\mathcal{B},\
\|\mathbf{x}_i-\mathbf{x}_{j^*(i)}\|_2\leq R,\\
0,
& \mathrm{otherwise}.
\end{cases}
\end{equation}

The $k$-th antibody binding arm is subsequently defined as the set of binder tokens assigned with class $k$:

\begin{equation}
\mathcal{B}_k=\{i\in\mathcal{B}|l_i^{B}=k\}.
\end{equation}

\subsection{Benchmark Evaluation}

\subsubsection{Nanobody benchmark evaluation}

For each target, each model generated 100 candidate structures, with 25 structures sampled from each template. Antibody regions were defined according to IMGT numbering~\cite{lefranc2003imgt}. Framework regions were kept fixed, and only CDRH1--3 regions were redesigned. With the framework sequence fixed, ProteinMPNN~\cite{dauparas2022robust} was used to generate eight candidate sequences for each structure. These sequences were internally ranked using the same BoltzGen-based ranking strategy, and the top-ranked sequence was selected for subsequent evaluation.

Candidate designs were ranked using the worst-ranking aggregation strategy introduced by BoltzGen~\cite{stark2025boltzgen}. Each design was independently ranked according to multiple evaluation metrics. Hydrogen bonds and salt bridges were identified using PLIP~\cite{Schake2025PLIP}. The metrics iPTM and minimum interface PAE were assigned a weight of 1, whereas binder pTM, hydrogen bonds, salt bridges and $\Delta$SASA were assigned a weight of 2. Let $r_{i,j}$ denote the rank of the $i$-th design under metric $j$. The final score was defined as:

\begin{equation}
s_i=
\max
\left\{
\frac{r_{i,\mathrm{iptm}}}{1},
\frac{r_{i,\mathrm{min\_ipae}}}{1},
\frac{r_{i,\mathrm{pTM}_{binder}}}{2},
\frac{r_{i,\mathrm{hbond}}}{2},
\frac{r_{i,\mathrm{saltbridge}}}{2},
\frac{r_{i,\Delta\mathrm{SASA}}}{2}
\right\}.
\end{equation}

The score of each design was determined by the maximum normalized rank among all evaluated metrics, corresponding to its worst-performing metric.

Geometric validity was evaluated using chain break and steric clash criteria. A design passed geometric filtering only when both criteria were satisfied. A chain break was defined as missing backbone atoms (C, N or CA) between adjacent residues within a binder chain, or when adjacent backbone distances exceeded the following thresholds:

\begin{equation}
d_{C-N}>2.0~\text{\AA}
\quad \mathrm{or}\quad
d_{CA-CA}>4.5~\text{\AA}.
\end{equation}

A steric clash was defined as any atom pair satisfying:

\begin{equation}
d_{\mathrm{atom-atom}}<1.1~\text{\AA}.
\end{equation}

Atom pairs within the same residue or between adjacent residues on the same chain were excluded, and covalent bond topology was not considered during clash detection.

\subsubsection{Bispecific antibody benchmark evaluation}

For the bispecific antibody benchmark, the CODV and diabody formats were derived from PDB 6O8D~\cite{wu2020trispecific} and PDB 9IP8~\cite{sato2025bispecific}, respectively. The tandem VHH format was constructed using two VHH heads derived from PDB 4KRL and PDB 4KRO~\cite{schmitz2013structural}, connected by a $(\mathrm{GGGGS})_4$ linker in the order of 4KRL-linker-4KRO.

Antibody variable regions were numbered according to IMGT. Framework region sequences and structures were fixed, while CDR regions were redesigned with fixed lengths according to the corresponding antibody format. If a linker region was provided in the crystal structure, the linker was not redesigned.

For each target, OFAntibody was used to sample 100 structures. ProteinMPNN was used to generate candidate sequences for each structure, with 8 sequences generated per structure. Rosetta~\cite{leaver2011rosetta3,alford2017rosetta} was then used for side-chain packing and constraint relaxation (Table~2). For each binder, the ProteinMPNN-generated sequence with the lowest Rosetta interface energy was independently selected for evaluation. A design was considered to satisfy the Energy Pass criterion when both binding arms exhibited interface energies lower than or equal to those of the corresponding reference complexes.

\begin{table}[htbp]
\centering
\caption{Parameter settings for Rosetta energy relaxation.}
\label{tab:rosetta_relax}
\begin{tabular}{ll|ll}
\toprule
\textbf{Arguments} & \textbf{Value} & \textbf{Arguments} & \textbf{Value} \\
\midrule
repeats & 1 & coordinate\_constraint\_weight & 0.5 \\
max\_iter & 50 & constraint\_atoms & CA \\
cartesian & TRUE & constraint\_sd & 1.0~\AA \\
relax\_scope & all & base scorefxn & get\_fa\_scorefxn() \\
allow jump & false & cart\_bonded & 0.5 \\
repack\_scope & design & pro\_close & 0 \\
\bottomrule
\end{tabular}
\end{table}

Hotspot Pass Rate and CDR Interaction Pass Rate were calculated based on sampled structures. Tandem VHH designs were evaluated using nanobody criteria, whereas CODV and diabody designs were evaluated using antibody criteria.

Hotspot Pass Rate was defined based on predefined hotspot constraints. A hotspot residue was considered satisfied when the minimum distance between its C$\alpha$ atom and any atom of the corresponding binding arm was below 9~\AA. Each binding interface was required to achieve hotspot coverage greater than 60\%, and all binding arms were required to satisfy this criterion.

CDR Interaction Pass Rate was defined based on format-specific CDR--antigen distance criteria. All CDR interaction rules for all binding arms were required to be simultaneously satisfied. The detailed CDR interaction criteria are summarized in Table~\ref{tab:cdr_interaction_criteria}.

\begin{table}[htbp]
\centering
\caption{CDR interaction criteria used for bispecific antibody design evaluation.}
\label{tab:cdr_interaction_criteria}
\begin{tabular}{lll}
\toprule
Format & CDR region & Interaction criterion \\
\midrule
\multirow{3}{*}{Tandem VHH}
& HCDR1 & $\geq$1 residue within 10~\AA \\
& HCDR2 & $\geq$1 residue within 12~\AA \\
& HCDR3 & $\geq$3 residues within 8~\AA \\
\midrule
\multirow{6}{*}{Diabody / CODV}
& HCDR1 & $\geq$1 residue within 15~\AA \\
& HCDR2 & $\geq$1 residue within 15~\AA \\
& HCDR3 & $\geq$3 residues within 8~\AA \\
& LCDR1 & $\geq$1 residue within 15~\AA \\
& LCDR2 & $\geq$1 residue within 15~\AA \\
& LCDR3 & $\geq$1 residue within 12~\AA \\
\bottomrule
\end{tabular}
\end{table}

Energy Pass Rate was determined by comparing the Rosetta interface energy of each binding arm with the corresponding reference complex. A binding arm was considered energy-pass when its interface energy was lower than or equal to the reference energy. A bispecific antibody design was considered energy-pass only when both binding arms satisfied this criterion.

\subsubsection{Joint design and independent binder assembly evaluation}

To mimic a modular assembly strategy, two monospecific VHH binders targeting the NM02 and 2Rs15d HER2 epitopes~\cite{sawmynaden2023co,d2017131i} were independently designed and subsequently assembled into a tandem VHH format. Candidate VHH binders were selected using single-VHH confidence criteria, including iPTM $>0.8$, minimum inter-chain PAE (min-iPAE) $<2.0$, and VHH chain pTM $>0.8$. The selected NM02 and 2Rs15d VHH candidates were then paired through exhaustive combination.

For each VHH pair, the two VHH structures were aligned by Kabsch~\cite{kabsch1976solution} rigid-body superposition using the shared HER2 chain A as the reference. The NM02 VHH was transformed into the coordinate system of the 2Rs15d VHH. Pairwise distances between non-hydrogen atoms of the two VHH domains were calculated after alignment. Combinations containing any inter-domain atom pair with a distance smaller than 2.0~\AA{} were considered sterically incompatible and removed.

For clash-free combinations, both possible VHH arrangements (2Rs15d$\rightarrow$NM02 and NM02$\rightarrow$2Rs15d) were evaluated based on the terminal C$\alpha$ distance. An arrangement was considered compatible with linker connection when the terminal C$\alpha$ distance was $\leq54.15$~\AA{}. Among compatible arrangements, the orientation with the shorter terminal distance was selected. A 20-residue Gly/Ser linker $(\mathrm{GGGGS})_4$ was then inserted between the two VHH domains according to the selected arrangement.

Linker conformations were generated using Rosetta GenKIC~\cite{bhardwaj2016accurate}. Structural relaxation was performed with coordinate constraints applied to the VHH head regions while allowing flexibility in the linker region. The resulting assemblies were evaluated using two-interface binding energy, linker energy, the repulsive energy term ($f_\mathrm{rep}$), and multi-conformation statistics.

The independent assembly procedure optimized only linker conformations and global geometric compatibility, while the amino acid sequences of the two VHH binders were kept unchanged.

\section*{Contributions}
Odin Zhang conceived the study.
Qinghan Wang and Zhiyuan Chen designed and implemented the network.
Gang Du, Heqin Zhu, Weibo Zhao, Jianmin Wang, Yuntao Yu, Ziyi You and Lijun Lang designed and conducted
the experiments and analyzed the results.
Qinghan Wang drafted the manuscript.
All authors discussed the results and revised the manuscript.

\bibliographystyle{valhalla_references}
\bibliography{references}


\end{document}